# The FLASHForward Facility at DESY

A. Aschikhin[1], C. Behrens[1], S. Bohlen[1], J. Dale[1], N. Delbos[2], L. di Lucchio[1],
E. Elsen[1], J.-H. Erbe[1], M. Felber[1], B. Foster[3,4,*], L. Goldberg[1], J. Grebenyuk[1],
J.-N. Gruse[1], B. Hidding[3,5], Zhanghu Hu[1], S. Karstensen[1], A. Knetsch[3], O. Kononenko[1],
V. Libov[1], K. Ludwig[1], A. R. Maier[2], A. Martinez de la Ossa[3],
T. Mehrling[1], C. A. J. Palmer[1], F. Pannek[1], L. Schaper[1], H. Schlarb[1], B. Schmidt[1],
S. Schreiber[1], J.-P. Schwinkendorf[1], H. Steel[1,6], M. Streeter[1], G. Tauscher[1], V. Wacker[1],
S. Weichert[1], S. Wunderlich[1], J. Zemella[1], J. Osterhoff[1]

[1] Deutsches Elektronen-Synchrotron (DESY), Notkestrasse 85, 22607 Hamburg, Germany
[2] Center for Free-Electron Laser Science & Department of Physics, University of Hamburg, Luruper Chaussee 149, 22761 Hamburg, Germany
[3] Department of Physics, University of Hamburg, Luruper Chaussee 149, 22761 Hamburg, Germany
[4] also at DESY and University of Oxford, UK
[5] also at University of Strathclyde, UK
[6] also at University of Sydney, Australia
[*] Corresponding author

## Abstract

The FLASHForward project at DESY is a pioneering plasma-wakefield acceleration experiment that aims to produce, in a few centimetres of ionised hydrogen, beams with energy of order GeV that are of quality sufficient to be used in a free-electron laser. The plasma wave will be driven by high-current density electron beams from the FLASH linear accelerator and will explore both external and internal witness-beam injection techniques. The plasma is created by ionising a gas in a gas cell with a multi-TW laser system, which can also be used to provide optical diagnostics of the plasma and electron beams due to the <30 fs synchronisation between the laser and the driving electron beam. The operation parameters of the experiment are discussed, as well as the scientific program.

## 1. Introduction

Interest in plasma wakefield-based particle acceleration techniques has greatly increased over the last decade, leading to substantial progress. Development has been such that most major large-scale particle-accelerator laboratories, such as BNL [1], CERN [2], DESY [3], LBNL [4], SLAC [5], have initiated or are successfully running their own research programs. In a plasma accelerator, a plasma wave is driven by either a co-propagating high-current-density particle beam [6,7] (PWFA, Plasma Wakefield Acceleration) or an intense laser pulse [8] (LWFA, Laser Wakefield Acceleration). The charge-density perturbation, established in the wake of the driver, results in strong electric fields from of the order of 10 GV/m up to hundreds of GV/m [9], depending on plasma and driver properties. These fields may be utilised to accelerate a trailing charged-particle beam (the witness bunch) to energies of order GeV over centimetre distances, thus promising a dramatic miniaturisation of particle acceleration modules when compared to standard radio-frequency technology. It is also possible to generate the witness beam by trapping plasma electrons into the accelerating phase of the wakefields. In this case, the witness-beam bunch length is a fraction of the plasma wavelength, $\lambda_p$, which itself is typically on the order of 10 to 100 µm, so that these beams intrinsically have bunch durations of a few femtoseconds. These properties make wakefield accelerators particularly attractive for photon-science applications in addition to interest in their use as compact accelerators for high-energy physics.

Due to the widespread availability and affordability of the laser systems required for laser-driven wakefield accelerators, the vast majority of experiments to date have utilized this approach to plasma wake generation. In contrast, beam-driven plasma wakefield accelerators have not yet received the same degree of investigation, largely due to the limited availability of particle accelerators with the requisite properties. Owing to the scarcity of research infrastructure, PWFAs trail behind LWFAs in their developmental progress, and hence, in demonstrated beam quality and applicability. Nevertheless, PWFAs possess a number of fundamental advantages over LWFAs: PWFA drivers propagate through plasma at close to the speed of light, *c*, naturally yielding a plasma-wave phase velocity much higher than in LWFA, where laser pulses propagate at their group velocity, $v_g < c$. Limiting effects such as overtaking and de-phasing of accelerated particles [10] are therefore mitigated [11]. In addition, strong transverse focusing fields in the plasma prevent driver expansion, allowing much longer acceleration lengths than possible in LWFA [12], with distances of order one metre already achieved [12]. The increased wake phase velocity also reduces unwanted self-injection of plasma electrons into the wakefield, therefore mitigating dark current (current due to un-wanted trapping), at the expense of making trapping of witness bunches more difficult. Furthermore, particle-beam drivers may provide greater parameter stability than high-intensity laser drivers, where observed shot-to-shot variations in the energy spectra of generated electron beams have been attributed in large part to fluctuations in the properties of the driving laser pulse in combination with nonlinear laser-pulse propagation effects in the plasma [13]. Another significant advantage of PWFAs over LWFAs for high-average-power applications is of a technical nature: particle beams for plasma wake excitation can be generated with Megawatt powers with efficiencies on the 10% level [14]. In contrast, state-of-the-art high-intensity laser systems deliver output powers of ~100 W with 0.1%-level wall-plug efficiency.

The FLASH accelerator at DESY [15] fulfills all critical technical requirements to support the delivery of suitable PWFA driver beams, and as such may be utilised to support many highly relevant studies in the field. The FLASHForward (Future ORiented Wakefield Accelerator Research and Development at FLASH) collaboration aims to construct a new electron beamline in the FLASH 2 tunnel, to demonstrate the potential of PWFAs for the production of high-quality electron beams that will support free-electron-laser (FEL) operation as a first step towards applications in high-energy physics.

In Section 2, the scientific goals for the FLASHForward project are outlined, with reference to specific milestones and the motivations behind each. Following this, Section 3 describes the technical design of FLASHForward, including the beamline extension to FLASH, and the adjoining laser and preparation laboratory. Important aspects of the PWFA process that will be demonstrated and developed, all crucial precursors to plasma accelerator application in high energy physics, are discussed in Section 4, before the current progress and future milestones of the project are outlined in Section 5.

## 2. Scientific Goals

The FLASHForward project has developed a broad scientific program to investigate a number of crucial issues that must be overcome, prior to the realisation of plasma-wakefield acceleration as a fully fledged disruptive future acceleration technique. Studies planned will address virtually all aspects of the PWFA process: from optimisation of driving-beam parameters to novel methods for witness-bunch generation, from plasma cell design and plasma shaping for greater stability to methods for improved release and capture of accelerated beams. With sufficient mastery of these techniques and careful characterisation of the accelerated electron beams, FLASHForward aims at demonstrating gain from a wakefield-driven free-electron laser for the first time. In addition, the lessons learned in the process will be invaluable for future staging of multiple reproducible plasma accelerators, potentially allowing energy boosting to the scales required for modern particle-physics research.

The specific scientific goals of FLASHForward are three-fold:

- Characterisation of externally injected electron beams from FLASH at an energy of around 1.25 GeV and their controlled release from a wakefield accelerator with energies > 1.6 GeV. Accordingly, the capabilities and performance of the FLASHForward beamline, from extraction at FLASH 2 to post-acceleration diagnostics, will be validated, and comparisons made with the results of extensive plasma cell [16] and beamline simulations performed to date [17].
- Exploration of novel in-plasma beam-generation techniques, to provide > 1.6 GeV energy, < 100 nm transverse normalized emittance, ~1 fs duration, and > 1 kA electron bunches. A number of such techniques, some proposed by members of the FLASHForward collaboration [18, 19, 20], can be implemented at FLASHForward to evaluate their effectiveness, as well as validating 3D particle-in-cell (PIC) simulations.
- Assessment of the potential of such beams for free-electron laser gain at wavelengths on the few-nanometre scale. That is, starting with a FLASH beam intended for FEL generation, boosting its energy via plasma acceleration and preserving its quality to observe subsequent FEL gain. Preservation of beam quality during acceleration, extraction from the plasma module, and transport to interaction or diagnostics regions is challenging [21]. It must be achieved as a prerequisite for staging of multiple plasma accelerators, essential for reaching energies at the frontier of particle physics.

## 3. FLASHForward Technical Design

The FLASHForward facility is designed to fulfill the scientific goals outlined in the previous section. It utilises the enabling features of the FLASH facility to exploit innovative concepts in plasma wakefield acceleration that make it unique in a number of aspects:

- FLASH provides world-leading possibilities for the synchronisation of electron beams with lasers on the few-10 fs level [22]. This is of paramount importance for laser-triggered witness-injection ("Trojan Horse") techniques [23,24]. These promise the generation of beams with unprecedentedly high brightness, which holds great promise for photon science, and low emittance without the necessity for damping rings, which will be important for high-energy physics applications. The demonstrated synchronisation accuracy is significantly better than at other existing facilities.

- FLASH is capable of producing beams with a double-bunch structure at the photocathode and, owing to the superconducting technology of the RF cavities, of accelerating and transporting them to the plasma cell without severe degradation from wakefield effects. The first bunch acts as a driver of the plasma wave while the trailing bunch forms a witness beam. The driver beam produced from the photo gun and transported to the plasma target has excellent beam properties, e.g. a ~2 µm normalized projected emittance in both transverse planes. This is a necessary requirement to achieve high-quality witness beams in such external injection techniques and, thus, essential for a careful study of the physics of staging. The excellent beam quality may be preserved when entering and exiting the plasma by deploying a novel windowless hydrogen target design that does not deteriorate beam quality by multiple scattering.
- The facility allows for unique techniques to shape the longitudinal phase-space of the driver beam, and by doing so, optimising the PWFA process. Shaping can be performed by usage of a third-harmonic RF cavity in combination with the FLASH bunch compressors to generate, e.g., triangular current profiles to give optimised transformer ratios [25]. In addition, FLASHForward features a variable R56 extraction dogleg plus collimator, allowing for post-compression of the FLASH beam to achieve multi-kiloamp currents for driving various internal injection techniques, some of which have been formulated by members of the FLASHForward team and of the Helmholtz Virtual Institute (VI) for Plasma-wave Acceleration [18][19][20][26], but not yet tested.
- An aspect particularly important for high-average-power applications is the FLASH capability to generate beams at repetition rates of MHz owing to its superconducting cavities. This will allow the stability of the plasma-acceleration process at µs time delay between two acceleration events to be tested, a crucial study for the future of PWFA in applications requiring high average power, such as high-energy physics.

Furthermore, the plasma cells recently developed by members of the FLASHForward collaboration and the VI will allow increasingly tunable plasma density profiles, improving controllability of the injection, acceleration, and release phases of PWFA.

### *3.1. FLASHForward Beamline*

The FLASH accelerator possesses a number of features that make it an ideal source for the FLASHForward beamline. In particular, the deliverable features of the plasma-wake driver beam include:

- A beam of sufficient quality to drive an FEL (1.25 GeV energy, ~0.1% energy spread, ~2 µm transverse normalised emittance) with 2.5 kA peak current
- Current profiles with variable length (10 to 500 fs) and flexible shapes (e.g. triangular [25])
- Laser-to-beam synchronization of < 30 fs rms [23]
- 10 Hz inter-bunch-train repetition rate, providing drive-bunch trains of up to 2 bunches with 1 $\mu s$ intra-bunch-train spacing

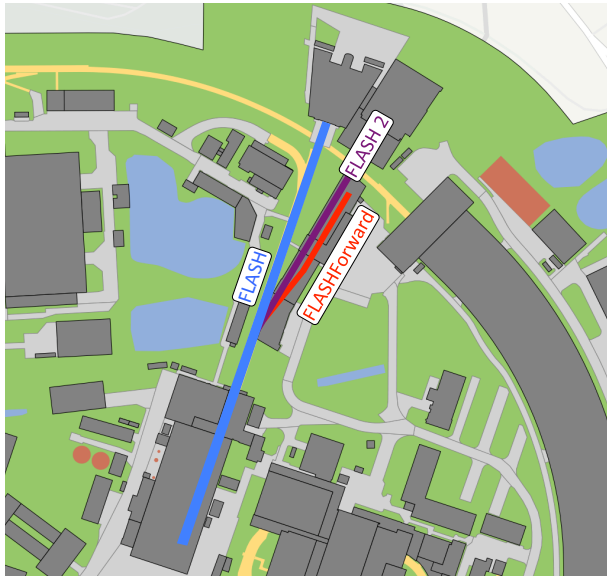

*Figure* 1–The FLASH accelerator complex on the DESY site including the new FLASH 2 buildings and the areas occupied by the FLASHForward beamline and infrastructure (shaded in red)

Following extraction from FLASH (*Figure 1*), the FLASHForward beamline, built within the FLASH 2 tunnel, consists conceptually of five sectors (*Figure 2*); beam extraction, beam matching and focusing, plasma cell, beam diagnostics, and the undulator section; in addition laser beams are transported into the facility.

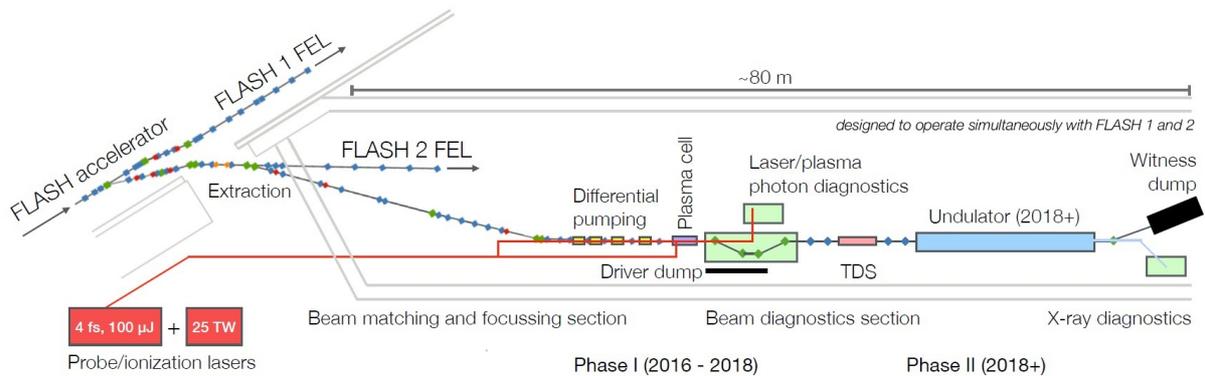

*Figure 2* – Configuration of the FLASHForward beamline with its main components, the electron beam extraction, beam matching and focusing, laser beamline, the central plasma interaction chamber, the beam diagnostics section, and the undulator section with photon diagnostics.

The extraction section is designed to achieve two goals: extraction of electron bunches from the FLASH 2 beamline, and forming the longitudinal current profile of the extracted beams by means of variable temporal compression. The FLASHForward and FLASH 2 beamlines operate simultaneously, with bunches separated spatially by two kicker magnets manufactured by the Budker Institute, Novosibirsk (each 25 cm long, with 8 winding turns and a 40 mm gap), which in combination provide an 8° kick at 1.6 GeV maximum energy. Each magnet's half-sine pulser (built by the DESY Machine Injection group) has design parameters of 2.24 kHz frequency, temporal jitter < 100 ns, B-field jitter of $10^{-4}$, and will allow the magnetic field (coupled to the electron beam inside a ceramic chamber) to switch polarity within 111.5 μs rise time. The resultant trajectory change is compensated by a set of dipole magnets (1×−0.8° and 2×−3.6°), giving a total transverse separation from the FLASH 2 beamline of 4 m. Inside this achromatic beam translation system, optics are chosen such that beam dispersion is closed up to second order, with $R_{16}, R_{166} \approx 0$ m (transverse displacement) and $R_{26}, R_{266} \approx 0$ (transverse angular displacement). High peak currents are achieved via temporal compression of the beam, possible by tuning the transport $R_{56}$ between -5 and 4 mm, and optimizing the RF-phase parameter. The beam energy distribution and current profile (*Fig. 3*) can be further improved by cutting the long tails, including the trailing small current bump. As the energy is correlated with the transverse offset in the dispersive section of the extraction beamline, it can be cut transversely with a collimator system, and since the beam is also energy-time correlated (chirped), the tails in the current profile are thus cut. The $R_{12}$ and $R_{22}$ values are tailored such that jitter specifications are fulfilled with $\frac{\Delta B}{B} \approx 10^{-4}$ kicker fluctuations.

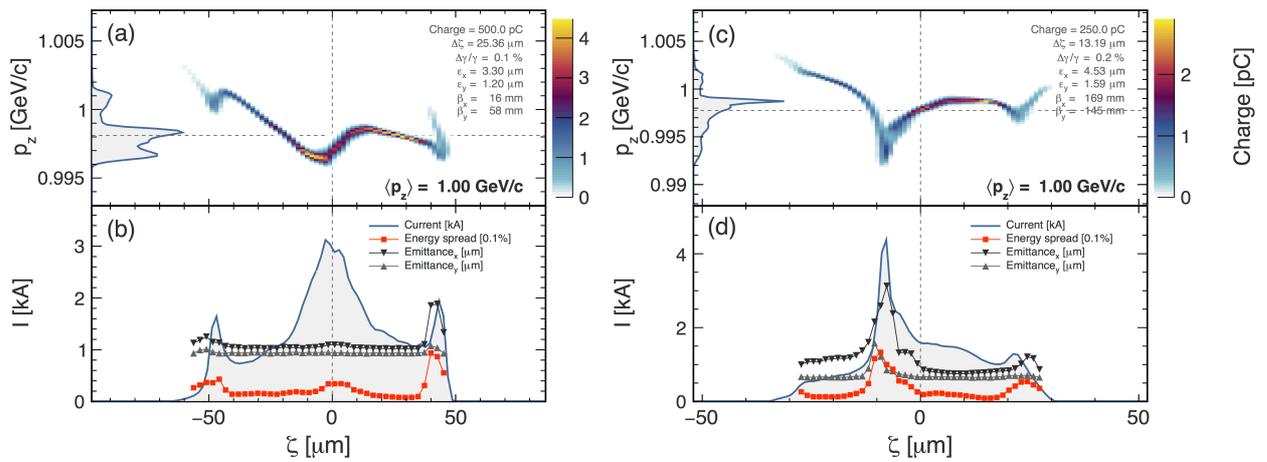

*Figure* 3 – Simulated example electron beams through the FLASHForward beam line at the entrance of the plasma target for charges of 500 pC and $R_{56}$ at its maximum (a,b), and 250 pC and $R_{56}$ at zero (c,d). The (a) and (c) panels show the longitudinal phase-space of the beams together with their projection on the $p_z$ axis. The (b) and (d) panels show the current profile of the beam (solid line), the sliced energy spread (squares), and the sliced normalized emittance in the horizontal (downwards triangles) and vertical (upwards triangles) planes. Units are as indicated in the panels.

Once parallel with the FLASH 2 beamline, the FLASHForward beam enters the focusing section. At its end the beam is spatially compressed to a transverse spot size of ~7 μm rms with minimal position (<10 μm) and angle (<0.5 mrad) jitter onto the plasma target in the interaction chamber. Also contained in the focusing section are three differential pumping stations to compensate for vacuum degradation caused by the gas loading of the plasma cell, in order to maintain the required $10^{-9}$ mbar at the beginning of the extraction section.

The central interaction chamber (Fig. 4) is the heart of the beamline, housing the interaction point of the ionising laser beam, electron beam, and plasma cell. It also contains various scintillator screens, transition-radiation screens, wires and knife edges, for use in the profiling and alignment of electron and laser beams. The central interaction chamber consists of a patented double vacuum-chamber design with one vacuum chamber on top of the other [27]. The experimental apparatus is mounted inside the upper vacuum chamber that adheres to the FLASH UHV cleanliness and vacuum requirements[1]. Hydrogen gas is used as the target in the plasma cell. The lower, less clean and higher pressure, vacuum chamber houses a hexapod positioning system to admit free translational and rotational control of the target, with its movement transferred to the experimental platform in the upper chamber by a bellows system. The chamber's side ports provide entry points for additional laser pulses for probing or particle-injection experiments.

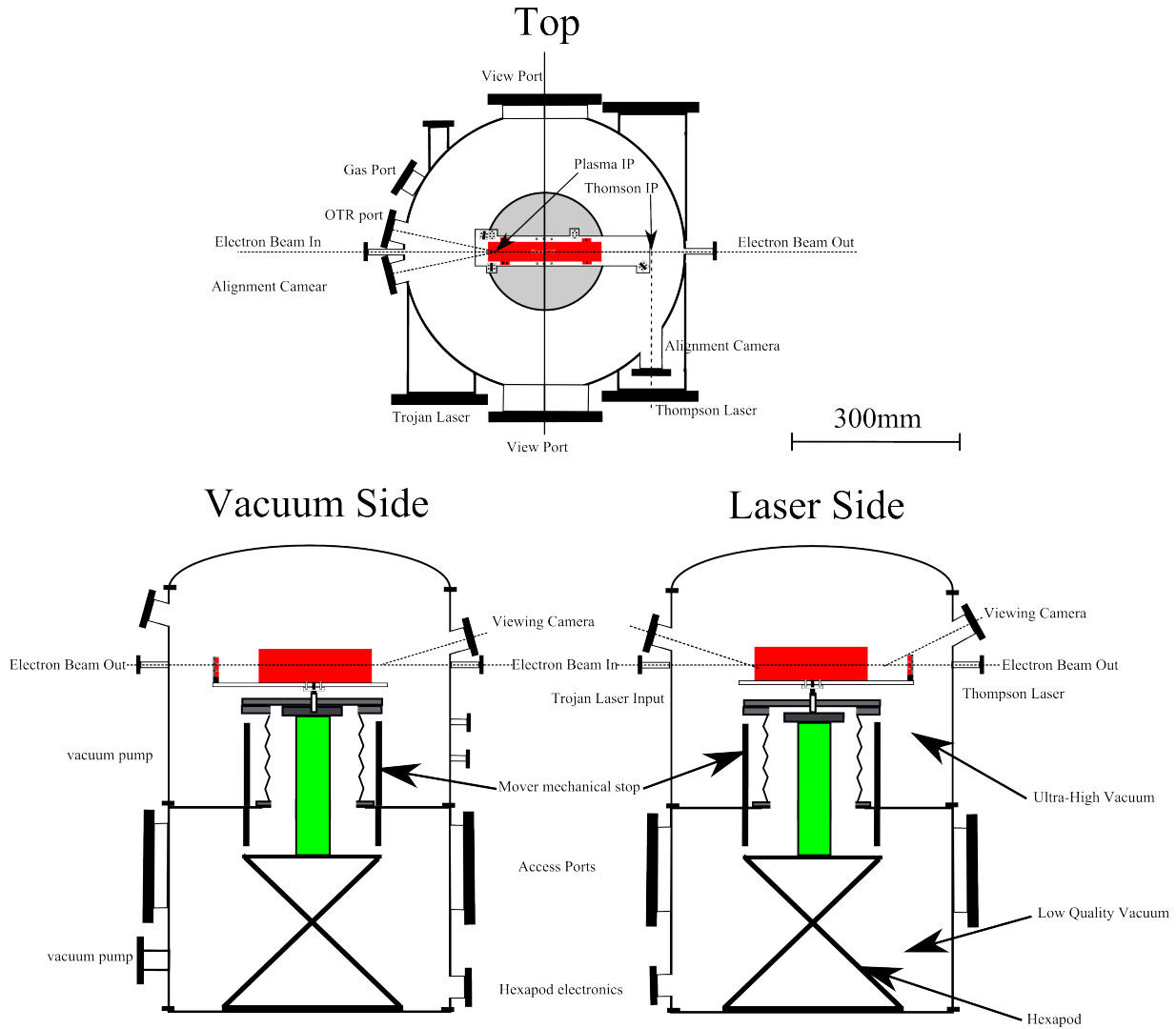

*Figure 4* – The FLASHForward central interaction chamber, which houses the PWFA cell. Side ports admit plasma gas lines, laser pulse inputs and cables to control a 6-DOF positioning system while alignment cameras operate via view ports.

The beam capture and diagnostics section (*Figure 5*) starts downstream of the interaction point, and is equipped to analyse the laser and driver/witness electrons following their interaction with the plasma. The post-plasma beamline is designed to transport and characterize electrons of up to 2.5 GeV. The initial section (beam capturing) consists of several quadrupole magnets focusing the divergent witness beam. After the quadrupoles, a dipole magnet acts as a broadband energy spectrometer. Several metres of drift space following the dipole are needed to reduce the size of the witness beam. A collimator is located at the point at which the beam is sufficiently small. It suppresses the drive beam while the witness beam passes through unaffected. The emittance-measurement section consists of another quadrupole magnet and a profiling screen. Finally, a high-resolution spectrometer precisely determines the energy distribution of the witness beam. Various diagnostics are located along the beamline, such as longitudinal profile measurement using transition radiation and diagnostics of the ionisation laser. In addition, this part of the beamline includes photon diagnostics to investigate the transmitted high-intensity laser radiation and the X-rays created inside the plasma by betatron radiation and by inverse Compton scattering of the laser off the electron beam. After passing through this beamline, the beam may be manipulated further to demonstrate lasing in the undulator.

---

[1] DESY technical specification No.: Vacuum 005/2008.

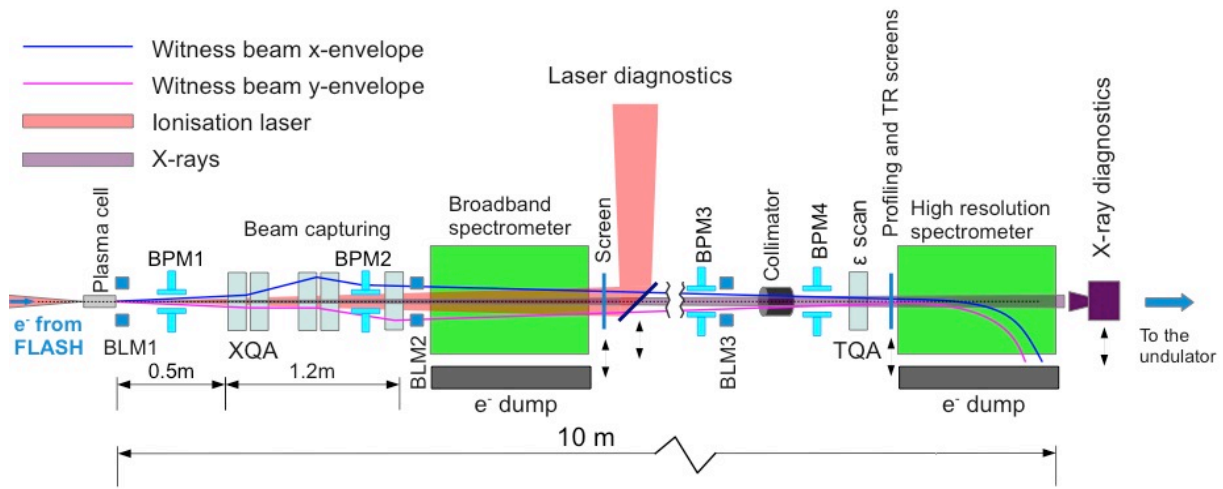

*Figure 5* – Post-plasma beamline layout: BPM stands for beam position monitors, BLM is the beam loss monitor while XQA/TQA denote electromagnetic quadrupoles.

### 3.2. Laser and Preparation Laboratories

The FLASHForward facility features a high-intensity laser system, utilised for purposes including the creation of the PWFA plasma medium, and as a diagnostic tool via direct interaction with the FLASH electron beam. This 25 TW – 10 Hz titanium-sapphire CPA laser system was built by Amplitude Technologies, France, and in testing has provided peak power of 26.0 TW, pulse durations of 24.5 fs (FWHM), pulse energy (after vacuum compression) of 633 mJ, and 1.25% RMS Energy stability (over 3 minutes). A secondary output of 3.5 mJ is compressed to 25 fs allowing for an independent probe beam. This beam can be coupled into a gas-filled hollow-core fiber to create spectral broadening via self-phase modulation and further compression down to ~5 fs with a chirped mirror compressor [28]

The timing instrumentation should achieve sub-30 fs synchronisation between laser pulses and the electron driver, critical for diagnostics and laser-triggered injection schemes. This is realised via a pulsed optical reference signal (rather than electronic RF signals) distributed throughout the FLASH facility, derived from the repetition rate of a femtosecond mode-locked laser oscillator [22], locked to the optical reference clock using feedback from a two-colour optical cross-correlator. A single-colour optical cross-correlator is used to measure changes in the round-trip time of the reference signal's fibre-optic network to a precision better than 10 fs [29], thus allowing compensation for the effects of environmental fluctuations via an adjustable optical delay. In addition, the relevant experimental areas are environmentally stabilised to an accuracy in temperature of 0.1 K and in humidity of 1%, thereby maintaining synchronisation jitter of laser pulses to the FLASH electron beam below a few 10 fs rms.

The BOND (Beam Optimisation and Novel Diagnostics) laboratory (*Figure 6*) is designed to allow focusing of test beams from the high-intensity laser inside the BOND experimental chamber (*Figure 6 (B)*). This chamber includes a 651 mm focal-length parabola to give laser focal intensity of 7.5 x $10^{18}$ W cm$^{-2}$, corresponding to a normalized vector potential ($a_0$) of 1.9, from which the beam is able to drive an LWFA cell setup in the quasi non-linear wakefield regime for plasma densities of 5 - 15 x $10^{18}$ cm$^{-3}$. The setup, contained in the ionisation test chamber (*Figure 6 (C)*), is used to mimic the ionisation and shaping of plasma targets in the central FLASHForward interaction chamber, permitting development and acceptance tests of equipment prior to installation in the FLASHForward beamline. The BOND laboratory is surrounded by metre-thick concrete shielding walls to provide protection against ionising radiation.

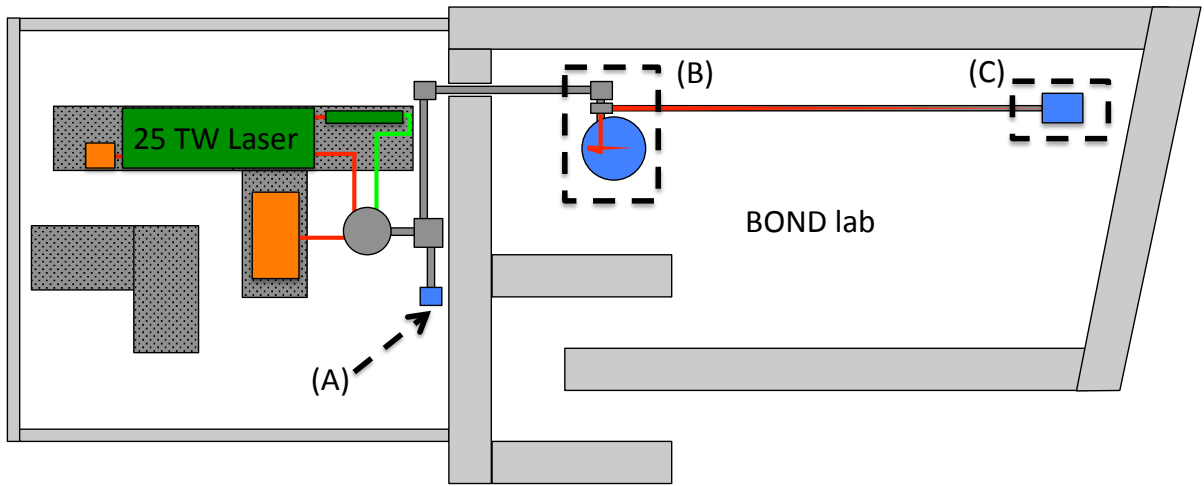

*Figure 6* – Laser and preparation laboratories for FLASHForward, consisting of laser lab, and BOND lab with 1 metre thickness concrete wall for shielding. (A) Laser output to FLASHForward beamline, (B) BOND experimental chamber (C) Ionisation test chamber.

## 4. Science Program

The FLASHForward beamline and adjoining laser and preparation laboratories allow continued study of a number of crucial aspects of the PWFA process that complement the ongoing numerical and theoretical work of the FLASHForward collaboration. These include (but are not limited to) novel in-plasma witness-bunch generation techniques, engineering and development of plasma-cell targets, and optimisation of the shape of the drive beam.

### 4.1. Witness-Bunch Generation

The method by which the witness bunch is injected dictates the emergent beam properties, including minimum transverse emittance, energy spread, and peak current [30] and is a limiting aspect of PWFA's effectiveness in accelerating such a bunch. The two broad approaches to injection, witness-bunch creation by either internal trapping of background plasma or ionised electrons or introduction of a pre-accelerated electron bunch, and different methods to achieve them, will be investigated in detail at FLASHForward.

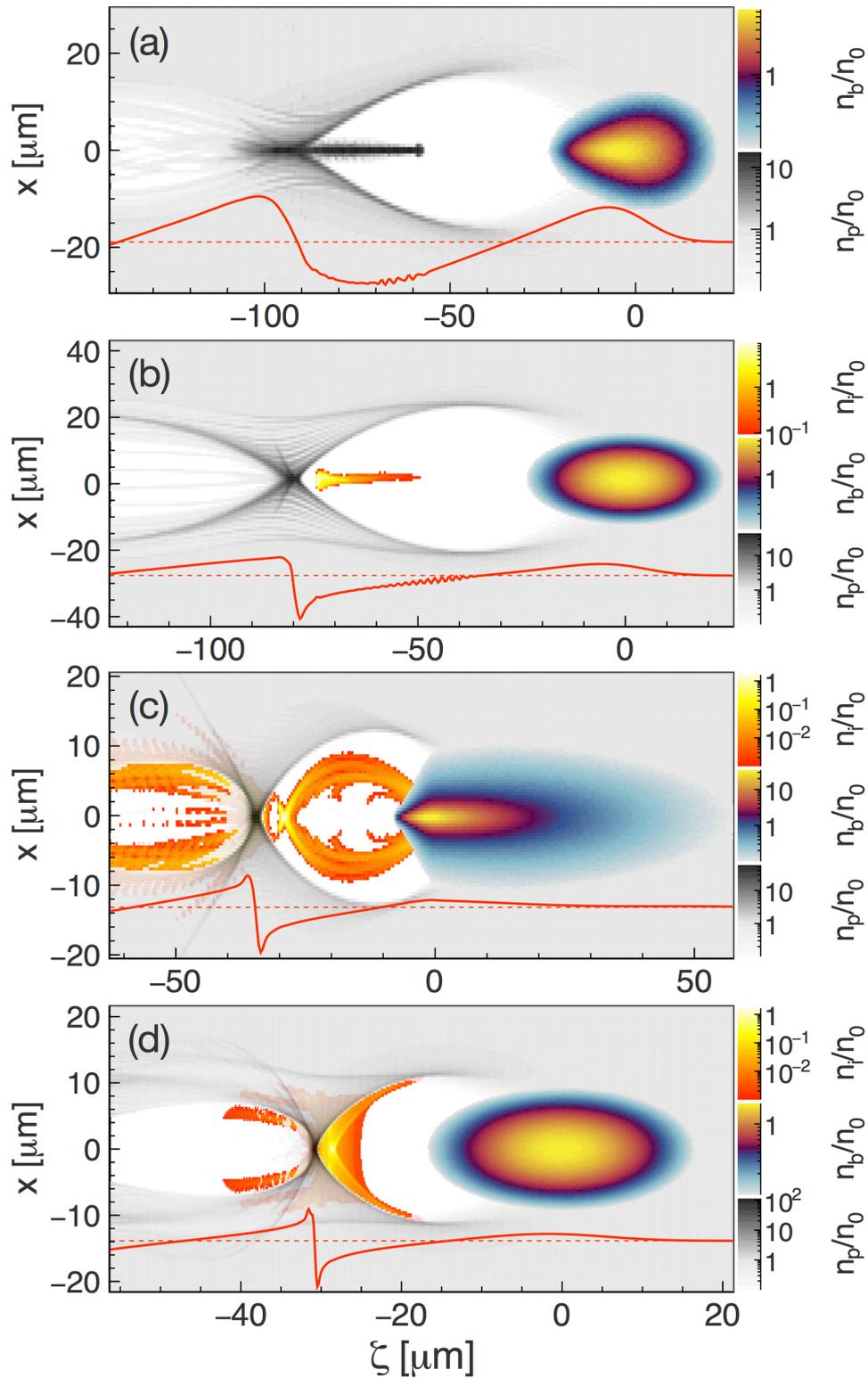

*Figure 7* – Charge densities of driving beam (red-blue palette), plasma electrons (black-white) and helium electrons (yellow-red), from 3D Particle-In-Cell (PIC) simulations of different internal witness-bunch generation techniques with the code OSIRIS [31,32,33]. a) Density down-ramp injection [34,35], b) Laser-induced ionisation injection [23], c) Beam-induced ionisation injection [18], d)

Wakefield-induced ionisation injection [19]. The red curves at the bottom of each sub-figure indicate the shape of the axial longitudinal electric field.

With sufficient control of wake-driver peak current and local plasma parameters, FLASHForward's technical capabilities allow access to a number of novel in-plasma beam-generation techniques (*Figure 7*):

- Density down-ramp injection (DDI), previously demonstrated in LWFA [30], requires plasma cells with adaptable plasma density profiles, which will be available at FLASHForward. Results of 3D OSIRIS PIC simulations (assuming the expected FLASHForward experimental conditions) have been promising, demonstrating narrow energy bandwidth (0.5% uncorrelated energy spread), high-current (0.7 kA), electron beams with small (0.2 μm) transverse normalised emittance [36]. The density down-ramp slope is found to strongly influence injected beam parameters, with less steep profiles inducing smaller charge and shorter bunches with larger transverse normalised emittance, making DDI an attractive injection method for PWFA owing to its tunability;
- Laser-induced ionisation injection ("Trojan Horse" injection) [23], which requires $I_B \gtrsim 5$ kA and a synchronised injection laser, used to release electrons from a second medium with higher ionisation threshold than the background plasma; the second medium is provided by a narrow high-density He / H gas jet at the entrance to the plasma cell;
- Beam-induced ionisation injection [18], which utilises radial electric fields of the driver to trigger ionisation, requires still higher beam currents ($I_B \gtrsim 7.5$ kA), as well as a secondary medium with higher ionisation threshold. Three-dimensional PIC Simulations with OSIRIS using realistic parameters for FLASHForward have yielded promising results with uncorrelated energy spread $\leq 1\%$, 3 kA peak currents, and ~1 μm normalised emittance [18];
- Wakefield-induced ionisation injection [19] uses only the wakefields to trigger ionisation and trapping of plasma electrons, and requires beam currents $I_B \gtrsim 10$ kA. Simulations assuming anticipated FLASHForward experimental conditions are promising, giving witness bunch parameters of ~2.5 GeV energy, uncorrelated energy spread < 1%, 5 kA peak currents, and ~ 0.3 μm transverse normalised emittance (*Figure 8*) [37]. This strategy for the injection is thought to provide improved control and stability over the accelerated bunches, as it is less sensitive to fluctuations in the driver's microstructure and does not rely on additional devices, such as lasers, for the injection.

It should be noted that the thresholds given above on the current of beam drivers to permit trapping in ionization-based techniques may be substantially reduced if assisted by density down-ramps that reduce the phase velocity of the plasma wake, as demonstrated by three-dimensional VSim [38] PIC simulations for the laser-induced ionisation injection case [20].

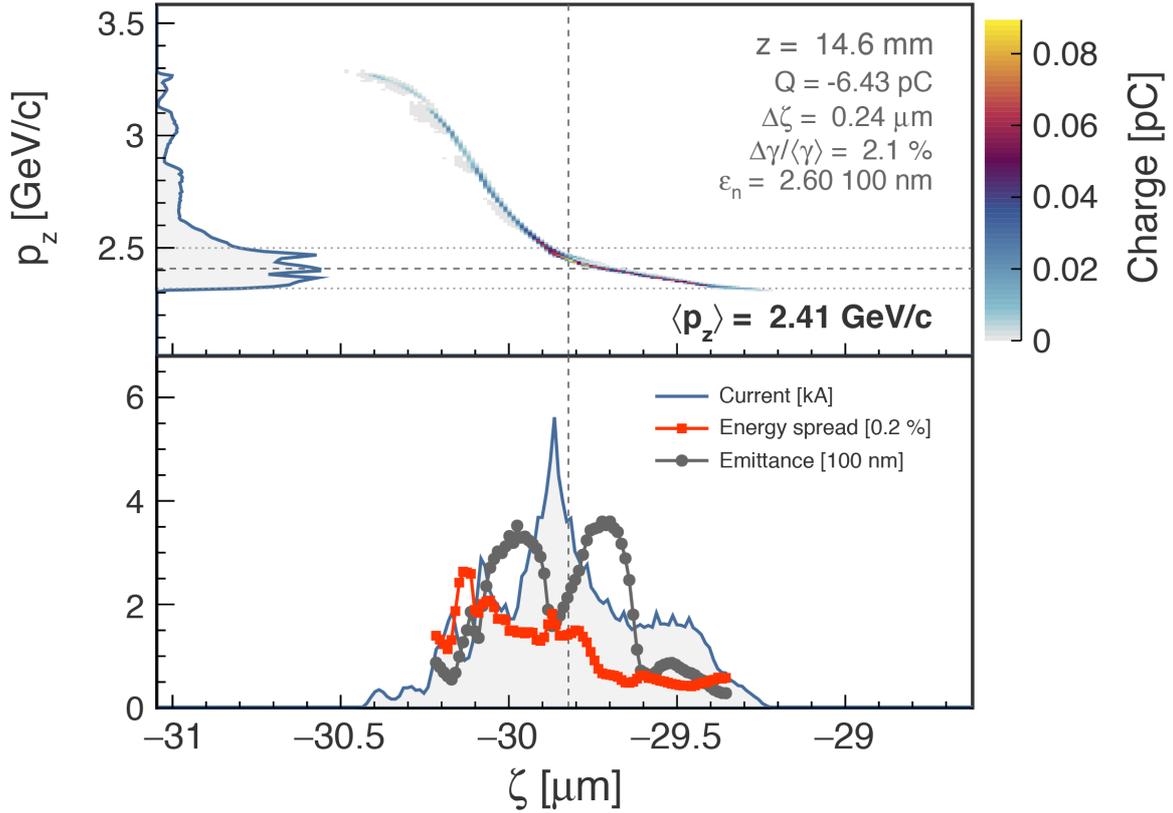

*Figure* 8 – Witness electron bunch injected in the plasma wake by means of wakefield-induced ionisation injection, after ~14 mm of acceleration in plasma. (a) Longitudinal phase-space of the beam together with its projection on the $p_z$ axis. (b) Current profile of the beam (solid line), sliced energy spread (squares), and sliced normalized emittance in the horizontal plane (dots). Units are as indicated in the panels.

Furthermore, methods for external witness-bunch generation have been simulated and tested at the FLASH facility. These methods offer the possibility of studying staging of plasma cells and may in addition allow for a higher degree of control over the witness-beam parameters. One such technique is generation of a pair of picosecond-separated bunches (a driver and a witness) at the FLASH photocathode, then transport through the accelerator within the same RF bucket (the same period of the RF accelerating fields). The two bunches are provided by a pulse split-and-delay system, which splits the initial laser beam using a polariser, where an adjustment in polarisation angle allows control of the relative intensity of the two beams. The separated beams are directed through optical paths of adjustable length, before being re-combined, yielding two pulses of nearly identical spatial characteristics (shape, transverse size, length) with variable charge and a variable delay between them, ready to pass through the accelerator. The plasma wavelength provides the driving constraint for this procedure; the initial laser pulse delay must be set such that post-acceleration compressed bunches have a spatial separation less than this wavelength, as required for wakefield acceleration in the same wake bucket.

In 2014 this process was tested at FLASH, with the aim of creating bunches of approximately equal intensity and initial photocathode separation of 20 mm (corresponding to ~ 65 ps delay). The pulse stacker operated effectively, generating the two bunches to within < 10 ps of the desired delay, and demonstrating production of double electron bunches at the gun without complications. Both bunches were then successfully transported through the accelerator in the same RF bucket; measurements using the existing FLASH transverse deflecting structure (TDS) showed two stably compressed and distinguishable resultant bunches, with 100 μm post-compression separation, and final energies of approximately 700 MeV and ±1% spread [17].

*4.2. Plasma-Target Development*

Precise and reliable tailoring of plasma-density profiles is identified as one of the critical points in achieving stable and reproducible conditions in plasma wakefield accelerators. At DESY, multiple input / output plasma cells have been designed, simulated in OpenFOAM [39] and fabricated via a combination of classical milling procedures on sapphire with femtosecond laser machining [16]. These cells are windowless (to reduce emittance growth), allow transverse laser probing (for characterisation of plasma densities and acceleration processes), and are compatible with plasma creation by ionisation laser, electric discharge, or electric fields from beams. Simulations demonstrate their capability to generate consistent electron densities along the beam's acceleration path, whilst maintaining a high-density He / H jet across the cell's entrance (as required for laser-induced (Trojan horse), Beam- and Wakefield-induced ionisation injection). These results are supported by the results of transverse Raman spectroscopy on prototype cells [16].

Whilst PWFA accelerated beams have, in general, small emittances, durations and transverse dimensions, this comes at the cost of a highly divergent beam upon exit from the plasma cell. Therefore, in order to stage multiple plasma acceleration cells, methods for overcoming the challenges of capture and transport of accelerated beams [21], whilst preserving their emittance, must be developed. One approach is a tailored plasma-density profile in the plasma-to-vacuum transition region subsequent to acceleration [40], necessitating testing and development of specialised plasma targets allowing the requisite level of control. Simulations using OpenFOAM have aided in the design of cells modified for this purpose, currently under construction, validating their capacity to form an adiabatic release region post acceleration. Simulations of the release region (*Figure* 9) demonstrate the merit of this approach, where increased plasma-density transition lengths yield marked reductions in emittance growth during beam extraction [40].

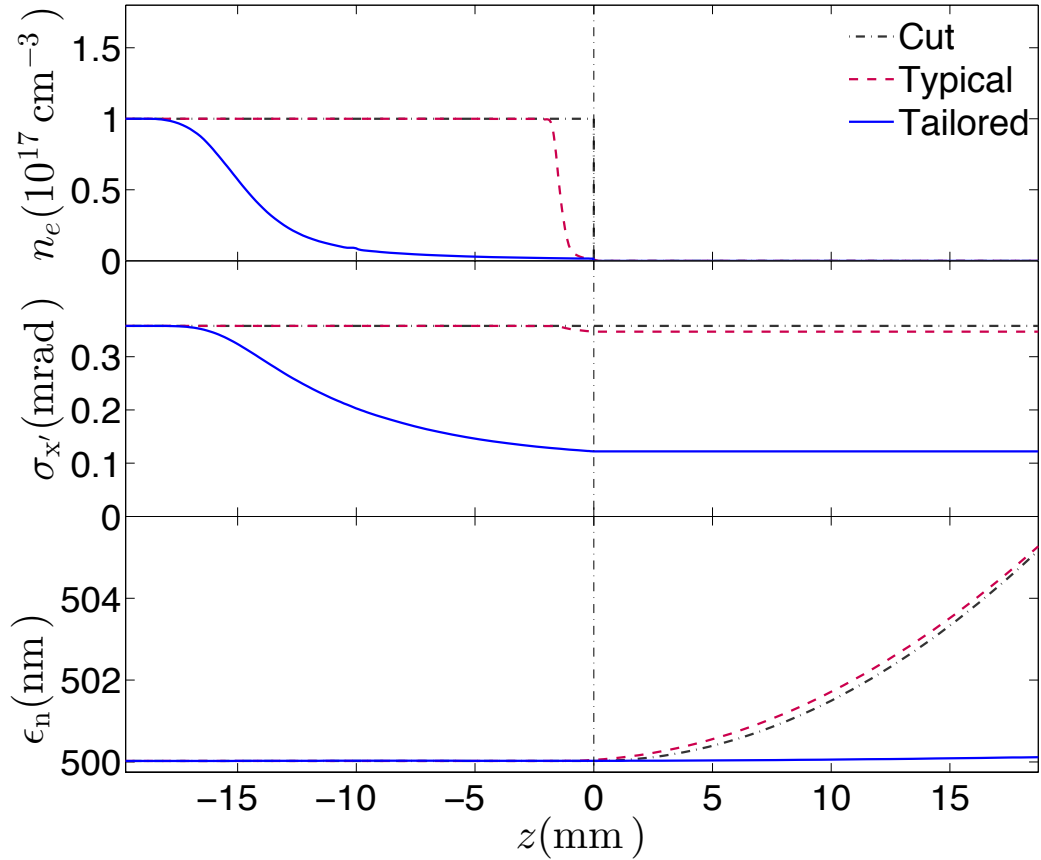

*Figure 9* – Simulated plasma-density profile (top), divergence (middle), and r.m.s. phase-space emittance (bottom) for a tailored density profile, a typical capillary gas-target outlet, and a density cut profile. The vertical dot-dashed line shows the location where plasma density has fallen to a negligible value. A substantial reduction in divergence and emittance is evident when a tailored plasma density profile (solid line) is used.

### *4.3. Driving beam-shape optimisation*

The limit for possible energy gain of a single electron accelerated by PWFA is given by the transformer ratio, which for a finite-length bunch with symmetric longitudinal charge distribution in the linear wakefield regime, can be proven to never exceed two [41,42]. Alternate approaches are thus desired, such as the use of triangular-shaped beams in the linear regime [42] and in the nonlinear regime [43]. FLASHForward aims to boost FLASH electrons to ~2 GeV (necessitating transformer ratios ~2, near the maximum for driving beams with Gaussian profiles), and by optimisation of the driving-beam parameters, boost particle energies to > 4 GeV, demonstrating transformer ratios substantially exceeding two. PIC simulations [44] predict that high-current drivers with symmetric profiles can generate transformer ratios ~3, when operating in a strong blowout regime, and triangular-shaped beams can reach even ~5. Tuning of the driving beam's current profile is possible in the extraction section of the beamline by means of a variable $R_{56}$. To this end, the question of which current profile should be selected to maximise the transformer ratio [45] will be the subject of future FLASHForward studies.

# 5. Current Status and Future Milestones

The FLASHForward project is separated into two phases, with each phase being divided into conceptual design phase, technical design phase, fabrication / procurement / installation phase, and the experimental campaign. At the present time phase 1 conceptual design of all beamline sections is completed, and technical design is well advanced, with installation of FLASHForward beamline components foreseen to finish in Q4 2016.

Phase 1 of the FLASHForward project will include construction and utilisation of most of the beamline (see *Figure 1* for the extent of Phase 1 design), as well as realisation of the first two science goals outlined in section 2. The installation period for the first phase began in Q2 2015, with commissioning to take place in Q1 2017. The first-round PWFA studies will then commence in Q2 2017 and continue for around 24 months, until the facility goes offline for installation of the Phase 2 beamline extension.

Significant scientific milestones for phase 1 include:

- Demonstration of density-down-ramp injection for PWFA – Q3/2017
- Demonstration of ionisation-based injection – Q4/2017
- Realisation of transformer ratios >2, exceeding the theoretical limit for Gaussian beams, using triangularly shaped driver beams – Q1/2018
- External witness-bunch injection and acceleration – Q3/2018
- Demonstration of laser-assisted ionisation injection for PWFA – Q1/2019

Simulation, fabrication, and testing of phase-2 apparatus will take place concurrently with the first phase experimental schedule, leading up to the beamline being taken temporarily offline in 2019 for installation of the phase-2 extension (*Figure 1*). This includes commissioning of the extended beamline before PWFA & FEL experimental studies resume. These phase-2 studies will address the third scientific goal outlined in section 2, FEL gain at few-nm wavelengths, and are scheduled to conclude in 2021.

# 6. Conclusions

The FLASHForward experiment promises to allow study of a range of techniques for, and aspects of, plasma-based acceleration. The FLASHForward beamline extends the existing FLASH accelerator facility, which has unique characteristics to provide driving beams ideal for PWFA methods. Following extraction into the FLASH-2 tunnel, the FLASHForward beamline manipulates incoming FLASH beams, allowing optimisation of driving-bunch parameters and access to a number of novel witness-bunch injection methods. A central interaction chamber houses a highly adjustable plasma acceleration cell, capable of providing tailored plasma-density profiles, as required for proposed techniques for combatting the tendency of the PWFA process to produce strongly diverging emergent beams. The subsequent beam capture and diagnostics section will aid in analysis of these methods, and for the second round of FLASHForward experimental study, will be extended to demonstrate FEL gain at few-nm wavelengths. With the conceptual design complete, and technical design well under way, FLASHForward is on-schedule for installation and to become fully operational in the first half of 2017.

# 7. Acknowledgements

We thank the management and technical support groups of the DESY Accelerator Division for their support; in particular we acknowledge the assistance of B. Faatz. We thank the OSIRIS consortium (IST/UCLA) for access to the OSIRIS code. Furthermore, we acknowledge the grant of computing time by the Jülich Supercomputing Centre on JUQUEEN under Project No. HHH23, at JUROPA under Project No. HHH20, at HLRN and at NERSC, and the use of thethe High-Performance Cluster (IT-HPC) at DESY. This work was funded by the Alexander von Humboldt Foundation through the award of a Humboldt Professorship to B. Foster, the Helmholtz Virtual Institute VH-VI-503 and the ARD program.

# References

[1] Accelerator Test Facility, http://www.bnl.gov/atf, Brookhaven National Laboratory, 2015, Accessed: 02/28/2015.

[2] Advanced Wakefield Experiment, http://awake.web.cern.ch/awake, CERN, 2013, Accessed: 02/28/2015.

[3] Laboratory for Laser- and beam-driven plasma acceleration, http://laola.desy.de, Deutsches Elektronen-Synchrotron, 2015, Accessed: 02/28/2015.

[4] Berkeley Lab Laser Accelerator, http://loasis.lbl.gov, University of California, 2014.

[5] FACET and Test Beam Facilities, http://facet.slac.stanford.edu, SLAC National Accelerator Laboratory, 2015.

[6] V. I. Veksler, Coherent Principle Of Acceleration Of Charged Particles. Proceedings of the CERN Symposium on High-Energy Accelerators and Pion Physics, (1956) 80–83, URL http://inspirehep.net/record/921325?ln=en.

[7] P. Chen et al., Acceleration of electrons by the interaction of a bunched electron beam with a plasma. Phys. Rev. Lett. 54 (1985) 693–696

[8] T. Tajima and J. Dawson, Laser electron accelerator, Phys. Rev. Lett. 43 (1979) 267.

[9] A. Modena et al., Electron acceleration from the breaking of relativistic plasma waves. Nature, 377 (1995) 606–608

[10] E. Esarey et al., Nonlinear Pump Depletion and Electron Dephasing in Laser Wakefield Accelerators. Advanced Accelerator Concepts: Eleventh Advanced Accelerator Concepts Workshop held 21-26 June, 2004 in Stony Brook, New York. Edited by Vitaly Yakimenko AIP Conference Proceedings, Melville, NY: American Institute of Physics, 737 (2004) 578-584

[11] A. Seryi et al., A concept of plasma wake field acceleration linear collider (PWFA-LC), No. SLAC-PUB-13766 (2009), http://www-public.slac.stanford.edu/sciDoc/docMeta.aspx?slacPubNumber=SLAC-PUB-13766

[12] I. Blumenfeld et al., Energy doubling of 42 GeV electrons in a metre-scale plasma wakefield accelerator, Nature, 445 (2007) 741-744.URL http://dx.doi.org/10.1038/nature05538

[13] P. Walker et al., Investigation of GeV-scale electron acceleration in a gas-filled capillary discharge waveguide, New Journal of Physics 15 (2013) 045024.

[14] E. Adli et al., Design of a TeV Beam Driven Plasma Wakefield Linear Collider, Proc. IPAC13, (2013), Ed. Zhimin Dai, C. Petit-Jean-Genaz, V.R.W. Schaa, Chuang Zhang, http://accelconf.web.cern.ch/AccelConf/IPAC2013/papers/tupme020.pdf.

[15] K. Tiedtke et al., The soft x-ray free-electron laser FLASH at DESY: beamlines, diagnostics and end-stations, New Journal of Physics 11 (2009) 023029.

[16] L. Schaper et al., Longitudinal gas-density profilometry for plasma-wakefield acceleration targets, Nuclear Instruments and Methods in Physics Research Section A: Accelerators, Spectrometers, Detectors and Associated Equipment 740 (2014) 208-211.

[17] C.M. Entrena Utrilla, Generation and transport of double-bunch electron beams in the FLASH beamline Internal Report DESY-2014-03169 (2014) (unpublished)

[18] A. Martinez de la Ossa et al., High-quality electron beams from field-induced ionization injection in the strong blow-out regime of beam-driven plasma accelerators, Nuclear Instruments and Methods in Physics Research Section A: Accelerators, Spectrometers, Detectors and Associated Equipment 740 (2014) 231-235.

[19] A. Martinez de la Ossa et al., High-Quality Electron Beams from Beam-Driven Plasma Accelerators by Wakefield-Induced Ionization Injection, Phys. Rev. Lett. 111 (2013) 245003.

[20] A. Knetsch et al., Downramp-assisted underdense photocathode electron bunch generation in plasma wakefield accelerators, arXiv preprint arXiv:1412.4844, (2014).

[21] T. Mehrling et al., Transverse emittance growth in staged laser-wakefield acceleration, Physical Review Special Topics - Accelerators and Beams 15 (2012) 111303.

[22] S. Schulz et al., Femtosecond all-optical synchronization of an X-ray free-electron laser, Nat Commun, 6 (2015), http://dx.doi.org/10.1038/ncomms6938

[23] B. Hidding et al. Ultracold Electron Bunch Generation via Plasma Photocathode Emission and Acceleration in a Beam-Driven Plasma Blowout . Phys. Rev. Lett. 108 (2012) 035001

[24] Y. Xi et al., Hybrid modeling of relativistic underdense plasma photocathode injectors, Physical Review Special Topics - Accelerators and Beams 16 (2013) 031303.

[25] P. Piot et al., Generation and Characterization of Electron Bunches with Ramped Current Profiles in a Dual-Frequency Superconducting Linear Accelerator, Phys. Rev. Lett. 108 (2012) 034801.

[26] Helmholtz VI for PWFA, https://vi-pwfa.desy.de/, Deutsches Elektronen-Synchrotron, (2015).

[27] J. Dale, K. Ludwig, L. Schaper, and J. Osterhoff, "Vorrichtung mit beweglicher Aufnahme für Vakuumkammern", Deutsche Patentanmeldung Nr. 10 2014 116 476.8, filing date November 11, 2014

[28] M. Nisoli et al., Generation of high energy 10 fs pulses by a new pulse compression technique. Applied Physics Letters 68 (1996) 2793-2795.

[29] J. Kim et al., Long-term femtosecond timing link stabilization using a single-crystal balanced cross correlator, Opt. Lett. 32 (2007) 1044-1046.

[30] C.G.R. Geddes et al., Plasma-Density-Gradient Injection of Low Absolute-Momentum-Spread Electron Bunches, Phys. Rev. Lett. 100 (2008) 215004.

[31] R. A. Fonseca et al. OSIRIS: A three-dimensional, fully relativistic particle in cell code for modelling plasma based accelerators. In P. Sloot, A. Hoekstra, C. Tan, and J. Dongarra, editors, Computational Science - ICCS 2002, volume 2331 of Lecture Notes in Computer Science, pages 342–351. Springer Berlin Heidelberg, 2002. ISBN 978-3-540-43594-5.

[32] R.A. Fonseca et al. One-to-one direct modelling of experiments and astrophysical scenarios: pushing the envelope on kinetic plasma simulations. Plasma Physics and Controlled Fusion 50 (2008) 124034.

[33] R.A. Fonseca et al., Exploiting multi-scale parallelism for large scale numerical modelling of laser wakefield accelerators. Plasma Physics and Controlled Fusion 55 (2013) 124011.

[34] S. Bulanov et al., Particle injection into the wave acceleration phase due to nonlinear wake wave breaking, Phys. Rev. E 58 (1998) 5257–5260.

[35] H. Suk et al., Plasma electron trapping and acceleration in a plasma wake field using a density transition. Phys. Rev. Lett. 86 (2001) 1011–1014.

[36] J. Grebenyuk et al., Beam-driven plasma-based acceleration of electrons with density down-ramp injection at FLASHForward, Nuclear Instruments and Methods in Physics Research Section A: Accelerators, Spectrometers, Detectors and Associated Equipment 740 (2014) 246-249.

[37] A. Martinez de la Ossa et al., Wakefield-Induced Ionization Injection in beam-driven plasma accelerators, submitted to Physics of Plasmas http://arxiv.org/abs/1506.05486.

[38] C. Nieter and J. R. Cary, VORPAL: a versatile plasma simulation code. J. Comput. Phys. 196 (2004) 448.

[39] OpenFoam Foundation, http://www.openfoam.com/.

[40] T. Mehrling, Theoretical and numerical studies on the transport of transverse beam quality in plasma-based accelerators. PhD Thesis. Universtat Hamburg, Institut fur Experimentalphysik (2014).

[41] K. L. F. Bane et al., Wake fields and wake field acceleration. AIP Conference Proceedings 127 (1985) 875–928. URL http://dx.doi.org/10.1063/1.35182.

[42] P. Chen et al., Energy transfer in the plasma wake-field accelerator. Phys. Rev. Lett. 56 (1986) 1252–1255. URL http://link.aps.org/doi/10.1103/PhysRevLett.56.1252.

[43] T. Katsouleas Physical mechanisms in the plasma wake-field accelerator. Phys. Rev. A 33 (1986) 2056–2064. URL http://link.aps.org/doi/10.1103/PhysRevA.33.2056.

[44] K.V. Lotov, Physics of Plasmas 12 (2005) 053105. URL http://dx.doi.org/10.1063/1.1889444.

[45] I. Blumenfeld et al., Scaling of the longitudinal electric field and transformer ratio in a nonlinear plasma wakefield accelerator, Physical Review Special Topics - Accelerators and Beams 13 (2010) 111301.